\documentstyle[aps,twocolumn,psfig,prb,floats]{revtex}

\begin{document}

\draft

\twocolumn[\hsize\textwidth\columnwidth\hsize\csname
@twocolumnfalse\endcsname

\title{Phonons and hole localization in 
La$_{1.475}$Nd$_{0.4}$Sr$_{0.125}$CuO$_{4}$}

\author{R. J. McQueeney\thanks{email: mcqueeney@lanl.gov}, J. L. 
Sarrao, J. S. Gardner, and M. F. Hundley}
\address{Los Alamos National Laboratory, Los Alamos, New Mexico 87545}

\author{R. Osborn}
\address{Argonne National Laboratory, Argonne, Illinois 60439}
\date{Received on \today}

\maketitle

\begin{abstract}
The phonon densities-of-states of 
La$_{1.475}$Nd$_{0.4}$Sr$_{0.125}$CuO$_{4}$ and 
La$_{1.9}$Sr$_{0.1}$CuO$_{4}$ were measured using inelastic neutron 
scattering.  The $\sim$70 meV phonon band which appears due to hole 
doping in La$_{1.9}$Sr$_{0.1}$CuO$_{4}$ is known to arise from strong 
electron-lattice coupling.  This phonon band is strongly suppressed in 
the Nd-doped compound for which $T_{c}$ is also suppressed, 
establishing a link between in-plane oxygen optical phonons and 
superconductivity.  This suppression is attributed to hole 
localization which becomes strong near the stripe ordering condition 
of certain cuprates near $x=1/8$ and competes with 
superconductivity.
\end{abstract}

]

The lattice dynamics of the high-temperature superconducting cuprates 
show evidence of very strong and unusual electron-lattice coupling.  
Early measurements of the phonon dispersion of 
La$_{2-x}$Sr$_{x}$CuO$_{4+\delta}$ and YBa$_{2}$Cu$_{3}$O$_{6+y}$ by 
Pintschovius, Reichardt, et al.\  show evidence for large softening 
(15-20 \% of the undoped phonon frequency) and broadening of
high-frequency phonon modes (in the range from 50-90 meV) as holes are 
doped into these compounds.
\cite{reichardt89,pintschovius91,pintschovius94,pintschovius96,reichardt96}
The softening appears to be directly proportional to the hole 
concentration, leading to the conclusion that the origin of these 
phenomena is the electron-lattice coupling.  These modes are 
associated with the oxygen half-breathing vibrations which propagate 
along the (1,0,0)-direction (Cu-O bond direction) in the CuO$_{2}$ 
plane.  It is puzzling that these modes should couple strongly to the 
holes because electronic structure calculations for La$_{2}$CuO$_{4}$ 
predict strong electron-phonon coupling along the (1,1,0)-direction
(i.e.\ the oxygen breathing modes), consistent with the nesting vector 
of the Fermi surface within the local-density approximation.\cite{krakauer}

Another testament to the size of the electron-lattice coupling, aside 
from the enormous frequency shifts, is the extent of the Brillouin 
zone which is affected.  Single-crystal phonon dispersion measurements 
show that the suppression of phonon frequencies occurs primarily in 
the region $0.25<q_{x}<0.5$, $q_{y}<0.15$ (in units of $2\pi/a)$, and 
to some unknown degree along $q_{z}$ for 
La$_{1.85}$Sr$_{0.15}$CuO$_{4}$.\cite{mcqueeney}  However, more 
convincing evidence of the extent of these interactions in q-space is 
the fact that they are easily observed in phonon density-of-states 
(DOS) measurements as the formation of a softer subband of phonon 
modes from the main band of in-plane oxygen modes that accounts for 
perhaps 10\% of the zone.  An example of this subband is shown by 
Renker et al.\ as a peak at $\sim$70 meV for 
La$_{1.85}$Sr$_{0.15}$CuO$_{4}$ which is known to grow at the expense 
of the 85 meV band as holes are added.\cite{renker87,renker92}  
A combination of all single-crystal phonon dispersion and 
polycrystalline phonon density-of-states measurements leads to a 
compendium of superconductors, with various crystal structures and 
methods of introducing holes, that display the similar phonon 
renormalizations, demonstrating the ubiquitous nature of the 
electron-lattice coupling.

The microscopic origin of these phonon anomalies has remained a 
mystery, and our collaboration has begun to look more carefully at 
these modes\cite{mcqueeney} and also those of related systems such as 
the nickelates.\cite{mcqueeney2,mcqueeney3} 
La$_{2-x}$Sr$_{x}$NiO$_{4}$ is isostructural to the 
La$_{2}$CuO$_{4}$-based superconductors but remains an insulator up to 
very high doping ($x\sim1$).  Measurements of the DOS of 
La$_{2-x}$Sr$_{x}$NiO$_{4}$ show no phonon softening from $x=0$ to 
$x=1/8$, in contrast to the large 70 meV subband formation for 
La$_{2-x}$Sr$_{x}$CuO$_{4}$.  This leads to a hypothesis that the 
charge fluctuations present in the metallic cuprate are a necessary 
ingredient in the phonon anomalies.  In this article, we compare the 
phonon DOS of La$_{1.9}$Sr$_{0.1}$CuO$_{4}$ and 
La$_{1.475}$Nd$_{0.4}$Sr$_{0.125}$CuO$_{4}$.  The Nd-doped compound 
has a pronounced ``1/8 anomaly'' at $x=1/8$, where the superconducting 
transition temperature ($T_{c}$) \cite{crawford} and other transport 
properties \cite{nakamura,buchner,lang,sera,baberski} are suppressed 
and static ordered charge/spin structures have been 
observed.\cite{tranquada96,tranquada97}  We find that the 70 meV 
phonon band is suppressed in the Nd-doped cuprate, establishing a 
possible link between superconductivity and the oxygen optical phonons.  
Similarly to the nickelates, the suppression of the phonon subband in 
the Nd-doped compound can also be interpreted as arising from reduced 
charge fluctuations, in this case by a slowing down of the hole dynamics. 

\begin{figure}
\centerline{
\psfig{file=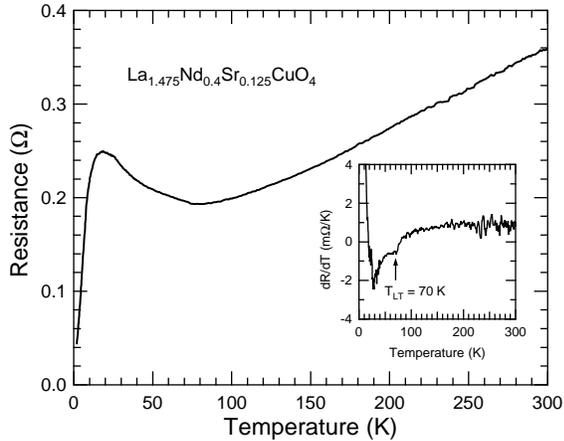,width=.45\textwidth}}
\caption{The temperature dependence of the electrical resistance of 
La$_{1.475}$Nd$_{0.4}$Sr$_{0.125}$CuO$_{4}$.  The inset figure shows 
the derivative of the resistance and indicates the LTO-LTT phase 
transition at T$_{LT}$=70 K.}
\label{fig1}
\end{figure}

Polycrystalline samples of La$_{1.475}$Nd$_{0.4}$Sr$_{0.125}$CuO$_{4}$ 
(LNSCO) and La$_{1.9}$Sr$_{0.1}$CuO$_{4}$ (LSCO) were 
prepared by standard solid state reaction of stoichiometric ratios of 
La$_{2}$O$_{3}$, Nd$_{2}$O$_{3}$, SrCO$_{3}$ and CuO.\@ Samples were 
subsequently annealed in flowing argon at 1100$^{\circ}$C for 20 
hours.  Samples weighed 60 and 90 grams, respectively.  The 
temperature-dependent resistivity of 
La$_{1.475}$Nd$_{0.4}$Sr$_{0.125}$CuO$_{4}$ was measured with a 
four-terminal low-frequency ac resistance bridge.  Sample contacts 
were made with a conductive silver epoxy.  The resistivity and its 
temperature derivative, shown in figure~\ref{fig1}, provide evidence 
for the LTO-LTT structural transition at 70 K and a broad 
superconducting transition with an onset at $\sim$15 K and midpoint at 
$\sim$7 K.\@ The broadness of the superconducting transition is 
presumably due to inhomogeneity in the large sample.

Time-of-flight inelastic neutron scattering measurements were 
performed on the Low Resolution Medium Energy Chopper Spectrometer 
(LRMECS) at Argonne National Laboratory's Intense Pulsed Neutron 
Source.  Scattering angles ($\phi$) from $1.95^{\circ}-120^{\circ}$ are 
covered by LRMECS.\@ For all measurements, an incident neutron energy of 
120 meV was chosen.  Powder samples were packed in flat-plate aluminum 
sample cans of dimensions 4"x3"x1/8" and oriented $45^{\circ}$ to the 
incident beam direction.  The sample can was mounted on a Displex 
closed-cycle He refrigerator for temperature dependence studies.  Each 
sample was measured at 10 K and 100 K.\@ In addition, empty aluminum can 
and white beam vanadium runs were performed for background and 
detector efficiency corrections, respectively.  The data were also 
corrected for a time-independent intensity arising from the ambient 
neutron background, the k'/k phase space factor and sample 
self-shielding.

To a first approximation the experimental intensity, 
$I(\phi,\omega)$, is proportional to the powder-averaged van Hove 
scattering function $S(\phi,\omega)$ plus a multiple scattering 
contribution $M(\phi,\omega)$, where $\hbar\omega$ is the energy 
transferred to the neutron.  The reduced data $I(\phi,\omega)$ were 
summed over all scattering angles for each data set, giving a function 
$I(\omega)$.  The multiple scattering was accounted for by 
extrapolating $I(\phi,\omega)$ to $\phi=0$ for several energy transfer 
ranges.  The intensity at the $\phi=0$ intercept is assumed to arise from 
multiple scattering.  The multiple scattering contribution is also 
assumed to be independent of $\phi$ (as warranted by the small 
self-shielding corrections) and is given by 
$M(\phi,\omega)=M(\omega)$. 

In order to determine the multiphonon scattering, the nuclear neutron 
scattering intensity of undoped La$_{2}$CuO$_{4}$ was calculated in 
the incoherent approximation after replacing the cross-section of La 
with the average (La,Nd,Sr) cross-section.  The partial phonon 
densities-of-states of La$_{2}$CuO$_{4}$ used in this calculation were 
obtained from a lattice dynamical shell model.  From $I(\omega)$, we 
subtract the multiple and multiphonon scattering intensities and also 
subtract an elastic gaussian which was obtained by fits to each data 
set.  Due to the encroachment of the elastic peak and the large error 
associated with its subtraction, only the portion of the DOS above 15 
meV can be reliably extracted.  The DOS is then obtained by 
multiplying by $\hbar\omega/(n(\omega)+1)$, where $n(\omega)$ is the Bose 
population factor.

\begin{figure}
\centerline{
\psfig{file=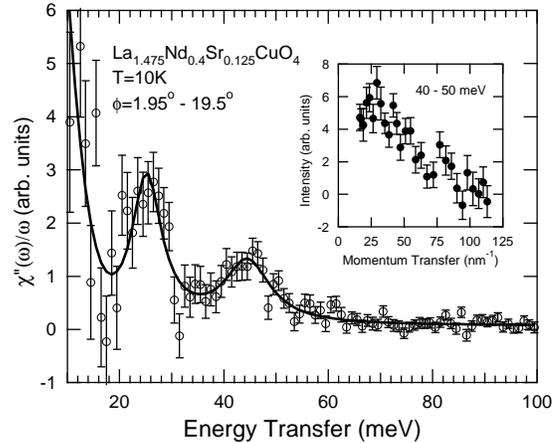,width=.45\textwidth}}
\caption{The difference of the inelastic neutron scattering spectra of 
La$_{1.475}$Nd$_{0.4}$Sr$_{0.125}$CuO$_{4}$ and 
La$_{1.9}$Sr$_{0.1}$CuO$_{4}$ at low angles and low temperature 
plotted as the dynamic magnetic susceptibility 
$\chi''(\omega)/\omega$.  Lorentzian fits to the spectrum (solid 
lines) give peak positions at 25.2 meV and 44.5 meV.\@ The inset shows 
that the intensity of the 44.5 meV excitation decreases with 
increasing momentum transfer, indicating a magnetic origin presumably 
arising from crystal-field excitations of the Nd$^{3+}$ ion.}
\label{fig2}
\end{figure}

\begin{figure}
\centerline{
\psfig{file=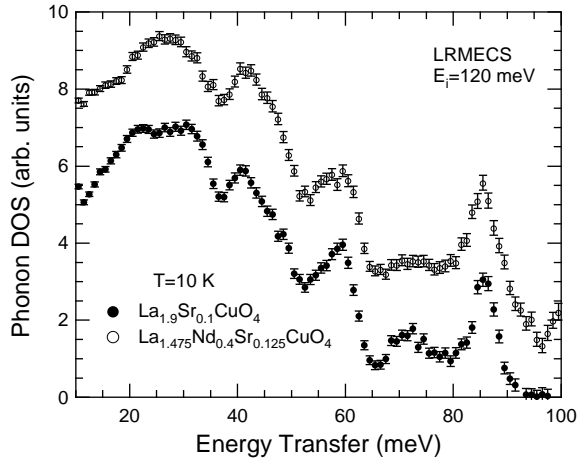,width=.45\textwidth}}
\caption{The generalized phonon densities-of-states of 
La$_{1.475}$Nd$_{0.4}$Sr$_{0.125}$CuO$_{4}$ and 
La$_{1.9}$Sr$_{0.1}$CuO$_{4}$ at 10K as obtained from inelastic 
neutron scattering spectra.  The areas at 25 and 45 meV for 
La$_{1.475}$Nd$_{0.4}$Sr$_{0.125}$CuO$_{4}$ have contributions from 
Nd$^{3+}$ crystal-field excitations which have not been subtracted 
out.}
\label{fig3}
\end{figure}

Absolute normalization of the DOS is made difficult by the presence of 
crystal-field excitations on the Nd-sites in 
La$_{1.475}$Nd$_{0.4}$Sr$_{0.125}$CuO$_{4}$.  These excitations were 
observed by comparison of the LNSCO and LSCO data sets summed over the 
low angle range from $1.95^{\circ}-19.5^{\circ}$.  Figure \ref{fig2} 
shows the difference of the low-angle averaged LNSCO and LSCO data 
sets at 10 K and plotted as the dynamic magnetic susceptibility 
$\chi''(\omega)/\omega$.  Lorentzian fits to the excitations, shown as 
a solid line in the figure, give peak energies of 25.2 and 44.5 meV.\@
These energy values are of the same order as the observed 
crystal-field splittings in Nd$_{2}$CuO$_{4}$.\cite{boothroyd92}\  The 
inset to fig. \ref{fig2} shows that the intensity of the 44.5 meV 
peak decreases with increasing momentum transfer, confirming the 
magnetic nature.  Even after summing over high angles the 
crystal-field intensity is still weakly present and appears as obvious 
but small differences between the LNSCO and LSCO data at 25 and 45 
meV. This is indicated in figure \ref{fig3}, which compares the DOS 
for each sample at 10 K. The LRMECS instrument has sufficient 
sensitivity to resolve such small relative changes in the data.

The phonon DOS for La$_{1.475}$Nd$_{0.4}$Sr$_{0.125}$CuO$_{4}$ 
and La$_{1.9}$Sr$_{0.1}$CuO$_{4}$ at $T=10 K$ are shown in figure
\ref{fig3}. The feature of the data we 
wish to discuss is the $\sim$70 meV phonon band which is not present 
in undoped La$_{2}$CuO$_{4}$.\cite{renker87,renker92}\  As discussed 
above, this band of modes softens out of the main 85 meV in-plane 
oxygen band with hole doping and is a signature of strong 
electron-lattice coupling in the cuprates.  This band appears 
prominently in the LSCO compound possessing a high T$_{c}$, but is 
surprisingly broadended and suppressed in LNSCO where T$_{c}$ is 
reduced.  By comparing these two materials, {\it we can associate the 
presence of the 70 meV phonon band with superconductivity}.

One may expect that the origin of the 70 meV band and its 
suppression could arise from local lattice vibrations and disorder 
associated with Sr$^{2+}$ or Nd$^{3+}$ dopants.  This is very unlikely 
for several reasons.  The 70 meV band does not form in 
La$_{1.85}$Sr$_{0.15}$NiO$_{4}$,\cite{mcqueeney2,mcqueeney3} which has 
the same crystal structure as LSCO, and is therefore unrelated to 
local structural effects or Madelung terms associated with Sr$^{2+}$ 
sites.  In general, these particular phonon renomalizations depend 
only on hole concentration and similar softening is observed after the 
addition of holes by excess oxygen in La$_{2}$CuO$_{4+\delta}$ 
\cite{pintschovius96,renker92} and YBa$_{2}$Cu$_{3}$O$_{6+y}$. 
\cite{reichardt89,reichardt96,renker92}\  The suppression in LNSCO 
does not originate from the increased disorder due to the Nd sites 
because such disorder would broaden many bands, especially 
(La,Nd,Sr)-O(2) vibrations at $\sim$55 meV.\@ With the exception of the 70 
meV oxygen phonon band, all other phonon bands are unchanged upon the 
addition of Nd.

It is worthwhile to discuss the suppression of the phonon anomaly in 
the Nd-doped cuprates in the context of electron-lattice coupling and 
the ``$x=1/8$ anomaly''.  It is well-known that superconductivity is 
destroyed for La$_{2}$CuO$_{4}$-based superconductors which have the 
low-temperature tetragonal (LTT) structure near hole concentrations 
$x=1/8$, as first detected in La$_{2-x}$Ba$_{x}$CuO$_{4}$.\cite{axe}\  
The La$_{2-x}$Sr$_{x}$CuO$_{4}$ superconductors exist in the 
low-temperature orthorhombic (LTO) structure.  The substitution of 
Nd$^{3+}$ or other rare earth ions for La$^{3+}$ in LSCO stabilizes 
the LTT phase without changing the hole concentration.cite{crawford}\  
In addition to the suppression of superconductivity, the LTT 
distortion strongly affects other transport properties, such as 
electrical resistivity,\cite{nakamura}\ 
thermopower,\cite{nakamura,lang}\ thermal 
conductivity,\cite{sera,baberski}\ and the Hall effect\cite{nakamura} 
and is a testament to the importance of the electron-lattice coupling 
in the cuprates.  In particular, the electrical resistivity has a 
semiconductor-like upturn below the LTO-LTT transition temperature 
($T_{LT}$) (see fig. \ref{fig1}) which signifies reduced mobility or a 
localization of the holes in the LTT phase.  B\"{u}chner et 
al.\cite{buchner} have shown that the upturn in resistivity and 
destruction of superconductivity in the LTT phase occur when the 
CuO$_{6}$ octahedral tilts responsible for the LTT phase (tilting 
around the [100] axis in tetragonal notation) exceed a critical angle.  
Tranquada et al.\cite{tranquada96,tranquada97} have used neutron 
diffraction to detect the presence of magnetic and charge superlattice 
peaks in the LTT phase of La$_{1.6-x}$Nd$_{0.4}$Sr$_{x}$CuO$_{4}$ 
which infer a static ordering of holes and Cu spins into a striped 
structure oriented along [100].  The octahedral tilts of the LTT phase 
correlate with the stripe direction and hole spacing for $x=1/8$ and 
commensurability effects between the two are hypothesized to pin the 
stripe structure.  In LSCO in the LTO phase, the LTO tilt pattern 
(tilt axis around [110]) is not commensurate with the stripe structure 
and the frustrated stripes are no longer pinned.  In superconducting 
LSCO however, dynamic stripe correlations persist.  The spatial 
correlations determined by the peak positions in LSCO are the same as 
in LNSCO, but become purely dynamic rather than elastic and signify a 
fluctuating stripe state.\cite{tranquada97,cheong91,yamada98}\
Tranquada et al.\ have asserted that superconductivity favors the 
presence of a dynamic stripe state and is in competition with the
order created by the pinning of stripe fluctuations by the LTT distortion.
The phonon behavior mimics this competition because the 70 meV phonon
band is most prominent in the superconducting compounds where purely 
dynamic stripe states are thought to exist. 

It is possible that reduced charge fluctuations in LNSCO near the 
stripe ordering temperature suppress the 70 meV phonon band.  Evidence 
for reduced charge fluctuations in LNSCO comes mainly from the 
aforementioned electrical resistivity measurements, neutron 
scattering, and also from far infra-red reflectivity measurements.  Obviously, 
the static stripe structures observed by neutron scattering imply 
reduced charge and spin fluctuations.  The downward shift of the 
magnetic spectral weight to the elastic region upon stripe 
ordering indicates a gradual freezing out of the holes and Cu 
spins.\cite{tranquada98}\  The infra-red results show that the optical 
conductivity is reduced in 
La$_{1.475}$Nd$_{0.4}$Sr$_{0.125}$CuO$_{4}$, as compared to 
La$_{1.9}$Sr$_{0.1}$CuO$_{4}$, over a wide energy range from $\sim$50 
- 500 meV with a corresponding decrease in spectral weight of
$\sim$10\%.\cite{tajima}\  How does this hole localization affect the phonon
spectrum?  In an exaggerated manner, we can compare the lattice 
dynamics of LSCO to La$_{2-x}$Sr$_{x}$NiO$_{4}$, which is 
isostructural to LSCO but remains an insulator up to $x \sim 1$.  We 
see no renormalizations of the oxygen phonons in 
La$_{2-x}$Sr$_{x}$NiO$_{4}$ at $x=1/8$.\cite{mcqueeney2,mcqueeney3}\
In reference to the insulating nickelates, the metallic nature of the 
cuprates and the presence of charge fluctuations appears to be a 
necessary ingredient in the phonon anomaly observed in high-T$_{c}$ 
compounds.  It seems reasonable to associate the suppression of the 70 
meV phonon band in LNSCO with slower hole dynamics.

One inconsistency in this interpretation is the temperature
independence of the phonon band in LNSCO above and below either the stripe 
ordering transition at $\sim$50~K \cite{tranquada96} or the LTO-LTT 
transition at 70~K.\@ The phonon DOS extracted for LSCO and LNSCO at
10 K 
and 100 K are unchanged. Of course, this does not imply 
that more subtle changes in the phonon dispersion, related to the 
stripe structure or the LTT distortion, cannot be observed in single-crystal 
measurements.  For example, the measured 
dispersion of the oxygen half-breathing branch in single-crystals of 
La$_{1.85}$Sr$_{0.15}$CuO$_{4}$ displays subtle temeprature dependent 
changes (discussed below), which are not apparent in polycrystalline 
DOS measurements.  However, it is evident that the gross differences 
in the phonon DOS of LNSCO and LSCO do not depend on the static stripe 
order of LNSCO or the long-range LTO/LTT crystal structures.

The lack of temperature dependence of the phonon anomaly is in strong 
contrast to static properties which change drastically at $T_{LT}$.\@  
The bulk transport properties of LSCO and LNSCO are similar above 
$T_{LT}$ and diverge below.  It would seem that the suppression of the 
70 meV band could not originate from reduced charge fluctuations, 
since holes are seemingly ``unpinned'' above $T_{LT}$.\@  However, there 
are several reasons to expect only small changes of the phonons at 
T$_{LT}$.\@  Hole pinning in the static stripe phase is certainly not 
complete since the upturn in resistivity in the stripe phase is rather 
modest and slow, incommensurate spin fluctuations exist both above and 
below the stripe ordering temperature in LNSCO.\cite{tranquada98}\@  
This implies that Nd-doping slows the dynamic stripe fluctuations and 
shifts spin and charge spectral weight down to low energies.  Bulk 
properties would only be sensitive to the static component below 
T$_{LT}$.\@  For the purposes of determining the effect of hole 
localization on the phonons, it is crucial to understand how the 
electronic susceptibility changes in an energy range closer to the 
oxygen phonon frequencies, rather than $\omega=0$.  We expect the 
electron-lattice interaction to be stronger when the electronic 
susceptibility has increasing weight at the phonon energy.  
It would therefore prove quite useful to 
determine the spin fluctuation spectrum of LNSCO up to $\sim$100 meV 
and compare this to LSCO.\@ While one can only infer the underlying 
charge dynamics from the spin fluctuation spectrum, slower stripe 
fluctuations would look more static to the phonon either above or below 
$T_{LT}$ and could suppress the phonon anomaly.  This 
may help to explain why the optical conductivity and the reduced 
spectral weight of LNSCO between 50 meV and 500 meV is also unchanged 
through $T_{LT}$.\cite{tajima}\  The finite energy spectroscopic 
results of both the phonon and charge system change very little 
through $T_{LT}$.

Another factor contributing to the insensitivity of the phonon anomaly 
to stripe order is its local behavior.  Previous inelastic neutron 
scattering work on single-crystals of 
La$_{1.85}$Sr$_{0.15}$CuO$_{4}$\cite{mcqueeney} has demonstrated that 
the phonon anomaly is related to, but different than, stripe order.  
The discontinuity of the Cu-O half-breathing phonon branch at ${\bf 
q}$=(0.25,0,0) is interpreted as an incipient doubling of the 
unit-cell ($\lambda=2a$), rather than a reflection of the stripe 
spacing of 4a along [100].  The phonon branch is flat between 
(0.25,0,0) and (0.5,0,0) indicating that these 
vibrations are more properly discussed in terms of localized oxygen 
modes near to localized holes and the electron-lattice coupling 
induces charge transfer between neighboring oxygens.  The phonon 
anomaly does not reflect the superlattice of the stripe structure 
unlike infra-red and Raman phonon signatures of stripe order in the 
nickelates.\cite{blumberg,yamamoto,katsufuji}\  Rather, the phonon 
anomaly depends on local charge fluctuations in the vicinity of the 
stripe and is independent of stripe order.  In the Nd-doped compounds 
above $T_{LT}$, such stripe correlations still remain 
dynamically.\cite{tranquada98}\  There is also evidence that, at least 
in La$_{2-x}$Ba$_{x}$CuO$_{4}$, LTT type octahedral tilts still exist 
above $T_{LT}$ and are correlated over $\sim$10 \AA.\cite{billinge}\

The phonon anomaly in high-T$_{c}$ cuprates seems to depend on 
charge fluctuations and their energy spectrum.  In order 
to completely characterize the electron-lattice interaction much more 
work is required.  Systematic studies of the single-crystal phonon 
dispersion of various compounds as a function of hole concentration 
seem necessary.  Given the relatively high energies and short
wavelengths of these phonons, only inelastic neutron experiments
requiring large single-crystals are sufficient at this time.  However, 
the phonons may prove to be as important as the spin fluctuations in 
probing the charge dynamics of the cuprates.  A large investment into 
their study may be worthwhile.

In summary, La$_{2-x}$Sr$_{x}$CuO$_{4}$ displays a large phonon
anomaly attributed to strong electron-lattice coupling and is indicated by the 
presence of a soft $\sim$70 meV oxygen phonon band in polycrystalline 
inelastic neutron scattering measurements.  The 70 meV phonon band is 
suppressed in La$_{1.475}$Nd$_{0.4}$Sr$_{0.125}$CuO$_{4}$, a cuprate 
with reduced $T_{c}$ due to hole order, demonstrating a link between 
superconductivity and the oxygen optical phonons.  By comparison of 
phonon densities-of-states in the cuprates to 
La$_{1.875}$Sr$_{0.125}$NiO$_{4}$, an insulator which has no phonon anomaly, we 
suggest that charge fluctuations are necessary in order to explain the 
phonon renormalizations.

\acknowledgments

RJM would like to thank T. Egami, A. R. Bishop, P. C. Hammel and J. M. Tranquada 
for useful discussions. This work is supported (in part) by the 
U.\ S.\ Department of Energy under contract No.  W-7405-Eng-36 with 
the University of California.  This work has benefited from the use of 
the Intense Pulsed Neutron Source at Argonne National Laboratory.  
This facility is funded by the U.\ S.\ Department of Energy, 
BES-Materials Science, uncer contract W-31-109-Eng-38.


\end{document}